\documentclass[aip,reprint]{revtex4-1}

\usepackage{graphicx}
\usepackage{color}
\usepackage{bm}
\usepackage{amsmath}
\usepackage{amssymb}
\usepackage{xspace}
\usepackage[normalem]{ulem}
\usepackage{units}
\usepackage{dcolumn}
\usepackage{paralist}
\usepackage{natmove}
\usepackage{natbib}
\usepackage{float}

\usepackage{hyperref}

\begin{document}

\title{Theoretical Prediction of Enhanced Thermopower in $n$-doped Si/Ge Superlattices using Effective Mass Approximation}
\author{Manoj~Settipalli}
\affiliation{Ann and H.J. Smead Aerospace Engineering Sciences, University of Colorado Boulder, Boulder, Colorado 80303, USA}
\author{Sanghamitra~Neogi}
\email{sanghamitra.neogi@colorado.edu}
\affiliation{Ann and H.J. Smead Aerospace Engineering Sciences, University of Colorado Boulder, Boulder, Colorado 80303, USA}

\begin{abstract}
We analyze the cross-plane miniband transport in $n$-doped [001] silicon (Si)/germanium (Ge) superlattices using an effective mass approximation (EMA) approach that correctly accounts for the indirect nature of the Si and Ge band gaps. Direct-gap based EMA has been employed so far to investigate the electronic properties of these superlattices, that does not accurately predict transport properties. We use the Boltzmann transport equation framework in combination with the EMA band analysis, and predict that significant improvement of the thermopower of $n$-doped Si/Ge superlattices can be achieved by controlling the lattice strain environment in these heterostructured materials. We illustrate that a remarkable degree of tunability in the Seebeck coefficient can be attained by growing the superlattices on various substrates, and varying the periods, and the compositions. Our calculations show up to $\sim3.2$-fold Seebeck enhancement in Si/Ge [001] superlattices over bulk silicon, in the high-doping regime, breaking the Pisarenko relation. The thermopower modulations lead to an increase of power factor by up to 20\%. Our approach is generically applicable to other superlattice systems, e.g., to investigate dimensional effects on electronic transport in two-dimensional nanowire and three dimensional nanodot superlattices. A material with high $S$ potentially improves the energy conversion efficiency of thermoelectric applications and additionally, is highly valuable in various Seebeck metrology techniques including thermal, flow, radiation, and chemical sensing applications. We anticipate that the ideas presented here will have a strong impact in controlling electronic transport in various thermoelectric, opto-electronic, and quantum-enhanced materials applications. 
\end{abstract}

\maketitle

\section{Introduction}
Dimensionally-confined semiconductor heterostructures, such as quantum wires~\cite{broido1995thermoelectric,lin2003thermoelectric,bertoni2000quantum}, superlattices~\cite{esaki1970superlattice,koga1999carrier,vashaee2004electronic,vashaee2006cross,vashaee2007thermionic,bahk2012seebeck,hinsche2012thermoelectric}, and quantum dots~\cite{van2002electron,kouwenhoven1997electron}, have been demonstrated to exhibit remarkable tunability of electronic and thermal transport properties, and thus offer great promises for energy transport applications. Consequently, heterostructured materials are actively being investigated to enable efficient and green energy transport and conversion applications, and to meet the increasing energy demands imposed by various modern day devices. The intriguing transport properties observed in these materials have been mainly attributed to the unique features in their density of energy states, distinguished from the bulk materials characteristics. Among various heterostructures studied, silicon (Si)/germanium (Ge) based heterostructures are of high technological relevance due to applications in the fields of electronics~\cite{thompson200490,meyerson1994high}, optoelectronics~\cite{koester2006germanium,liu2010ge,tsaur1994heterojunction,pearsall1994electronic,engvall1993electroluminescence}, thermoelectrics~\cite{chen2003recent,dresselhaus2007new,alam2013review}, and quantum materials~\cite{shi2011tunable} to name a few. With the advancements in nanofabrication techniques like molecular beam epitaxy (MBE)~\cite{kuan1991strain,david2018new} method, it is now possible to grow defect-free Si/Ge based heterostructures. Moreover, Si and Ge are non-toxic, cheap, readily available, and can be easily integrated into the current Si based technology making them strong candidate materials for a broad class of energy applications. Therefore, to address the ever-growing need for efficient and green energy conversion, it is imperative to acquire a fundamental understanding of the transport properties of Si/Ge heterostructures, particularly aimed at their thermoelectric applications. 

A thermoelectric material converts energy at its peak efficiency when the figure of merit, $ZT=S^2 \sigma T/\kappa$, is maximized, where $S$ is the thermopower or the Seebeck coefficient, $\sigma$ is the electrical conductivity, $\kappa$ is the total thermal conductivity, and $T$ is the absolute temperature. In the past decade, significant amount of research aimed at improving the $ZT$ by reducing $\kappa$~\cite{lee1997thermal,borca2000thermal,huxtable2002thermal,liu2003cross,bao2005electrical,alvarez2008cross,savic2013dimensionality}. It was observed that $\kappa$ of semiconductor superlattices is strongly reduced due to the presence of interface scattering mechanisms~\cite{chen1997thermal,koga2000experimental}. Therefore, it is desirable to improve the factor $S^2 \sigma$, known as the electronic power factor (PF), in order to further improve $ZT$. Enhancing $S$ has been reported to enhance the PF of semiconductor superlattices~\cite{vashaee2006cross,bastard1981superlattice}. In addtion to improving the energy conversion efficiency, a material with high $S$ is advantageous in various Seebeck metrology techniques including thermal, flow, radiation, and chemical sensing applications~\cite{van1986thermal,bakker2012nanoscale}. Therefore, it is beneficial to discover approaches to improve the thermopower and the PF of heterostructured materials for various technological applications. 

In a seminal paper, \citet{koga1999carrier} introduced the carrier pocket engineering (CPE) concept that the PF of the Si/Ge superlattices can be improved by varying the electronic properties of the well and the barrier regions, leading to significantly improved $ZT$s. $ZT=0.24$ and 0.96 were predicted for $n$-doped [001] and [111] Si/Ge superlattices at 300 K, respectively, compared to the bulk Si $ZT$ of 0.014 at 300 K, and encouraged researchers to explore the Si/Ge superlattices with more rigor~\cite{koga1999carrier}. The theoretical analysis in this study~\cite{koga1999carrier} was carried out employing a Kr\"{o}nig-Penney (KP) type model that ignored the indirect nature of the Si and Ge electronic band gaps. And no details were provided regarding the thermopower contributing to the improved $ZT$. A first principles density functional theory (DFT) study on strained [111] Si/Ge superlattices showed that $ZT$ can indeed be improved for these superlattices,  establishing the CPE concept~\cite{hinsche2012thermoelectric}. Interestingly, for $n$-doped [111] Si/Ge superlattices, no significant enhancements in $S$ were observed. Moreover, it was shown that $S$ very closely resembles the bulk Si behavior, following a Pisarenko-like relationship. On the other hand, our recent DFT study demonstrated that strain engineering the $n$-doped [001] Si/Ge superlattices can break the Pisarenko relation for $S$ in the cross-plane direction leading to significant increase of $S$~\cite{proshchenko2019optimization,proshchenko2019modulation}. However, the effect of the enhancements in $S$ on the PF has not been discussed. A complete understanding of the electronic transport properties of the highly technologically relevant [001] Si/Ge heterostructures is still missing.

In this article, we present our theoretical predictions of the cross-plane $S$ of $n$-doped [001] Si/Ge superlattices, along with its implications on the PF. We account for the indirect nature of the Si and Ge band gaps, by using an indirect-gap based effective mass approximation (EMA) model~\cite{bastard1981superlattice,trzeciakowski1988effective,rossi2011theory,mukherji1975band}, for the miniband dispersion of these superlattices, as opposed to the past KP model studies~\cite{zachai1990photoluminescence,koga1999carrier}. From the miniband dispersions obtained, we employ the Boltzmann transport equation framework with constant relaxation time to show that the cross-plane $S$ of various Si/Ge superlattices can be modulated varying strain, period and composition. We show that the Pisarenko relation for $S$ is broken not only by inducing substrate strain in the superlattices, but also by varying period and composition of strain-symmetrized superlattices. We predict up to $\sim 3.2$-fold enhancement of the cross-plane $S$ when compared to the bulk Si in the high-doping regime. Additionally, $S$ of these superlattices shows a non-monotonic behaviour with $T$ in the low-doping regime. This indicates that $ZT$ may not monotonically increase with $T$ as is expected for Si/Ge superlattices. The primary advantage of our EMA based approach is that it is much faster than other high accuracy methods, such as DFT, especially for larger systems, and it helps us to form preliminary intuition about the systems considered.

\section{Method}
We derive the analytical dispersion relations of the [001] Si/Ge superlattice energy bands employing the EMA, also known as the envelope function approximation~\cite{bastard1981superlattice,trzeciakowski1988effective,rossi2011theory}. In the EMA, the superlattice energy band dispersion relations are determined using the bulk parameters of its constituents and the band offsets. Previous studies that used EMA to compute energy bands of the Si/Ge superlattices used KP-like models that accounted for the correct superlattice band gap values, however, ignored the indirect nature of the Si and Ge band gaps~\cite{zachai1990photoluminescence,koga1999carrier}. Ignoring the indirect/multivalleyed nature of the CBM has a direct effect on the predicted superlattice transport properties. Here we follow the  multivalley band structure analysis method presented by \citet{mukherji1975band} in our EMA implementation to account for the indirect nature of the Si and Ge CBM. We then use the superlattice bands obtained with EMA to compute the electronic transport properties using Boltzmann transport equation (BTE) within the constant relaxation time approximation.

\subsection{Analytical Dispersion Relations of Si/Ge Superlattice Bands Using Effecting Mass Approximation}

\subsubsection{Conduction Bands}
It is well known that the 6-fold degenerate $\Delta$ valleys form the Si conduction band  minima (CBM), while the 8-fold degenerate $L$ valleys form the Ge CBM~\cite{ashcroftmermin}. However, the miniband energy levels from the  $L$ valley states are much higher than those from the $\Delta$ valleys in [001] Si/Ge superlattices~\cite{koga1999carrier}. As a consequence, the electronic transport in \textit{n}-doped [001] Si/Ge superlattices is dominated by the $\Delta$ valley states~\cite{koga1999carrier}. We derive the conduction miniband dispersion (CMB) of the $\Delta$ valley states using EMA. It is necessary to identify the lattice spacings and the superlattice potential profiles to obtain the miniband dispersion relations. In relaxed configurations, Si and Ge lattice constants are $a_{\text{Si}}=5.431\;\text{\AA}$ and $a_{\text{Ge}}=5.658\;\text{\AA}$, respectively~\cite{jones2002general}. In a strain-symmetrized (SS) superlattice, both the Si and Ge components are strained due to this lattice mismatch. Additionally, superlattices are usually grown on substrates which can induce further strain. We compute the lattice parameters of the SS and the substrate strained superlattices by using the macroscopic elastic energy minimization approach~\cite{van1986theoretical}. We assume that the in-plane lattice constants of both the constituents of a substrate strained superlattice are matched to the substrate lattice constant $a_{\parallel}$. The cross-plane lattice constants of the components are given by
$a_{i\perp}=a_i[1-D_{[001]}^i(a_{\parallel}/a_{i}-1)]$, where $a_i$ represents unstrained lattice constants with $i=$ Si, Ge, $D_{[001]}^\text{Si} = 0.776$, and $D_{[001]}^\text{Ge}=0.751$~\cite{van1989band}. The in-plane and cross-plane strain in the superlattices are defined as $\epsilon_{i\parallel}= (a_{\parallel}/a_{i}-1)$ and $\epsilon_{i\perp}= (a_{i\perp}/a_i-1)$, respectively~\cite{van1986theoretical}. In this article, we investigate strain-symmetrized superlattices with varied period and composition, and the superlattices grown on substrates that induce in-plane strains in Si and Ge, ranging from $\epsilon_{Si\parallel} = 0\%-4.2\%$ (tensile) and $\epsilon_{Ge\parallel} = 0\%-4\%$ (compressive), respectively~\cite{zachai1990photoluminescence,van1986theoretical}. We assume that the interface is smooth and the superlattice is periodic in the in-plane directions. 

The strain in the components splits the 6-fold degenerate $\Delta$ valleys of unstrained Si and Ge into 2-fold degenerate $\Delta_\perp$ valleys and 4-fold degenerate $\Delta_\parallel$ valleys. We compute the strain controlled potential profiles of the $\Delta_\perp$ and $\Delta_\parallel$ valleys in each material using the deformation potentials of bulk Si and bulk Ge\cite{van1986theoretical}. We denote the strain-split valley minima by $V^j_i$ with $i=\text{Si or Ge}$ and $j=$ $\perp$ or $\parallel$, respectively. The Si $\Delta$ valleys form the well regions and the Ge $\Delta$ valleys form the barrier regions in all the superlattices studied in this work~\cite{zachai1990photoluminescence}. We acknowledge that the two-fold degenerate $\Delta_\perp$ valleys are multi-valleyed and intervalley mixing effects can result in the lifting of degeneracy of these valleys~\cite{valavanis2007intervalley,ting1988valley}. However, these effects decrease with increasing period and are predicted to result in sub-band splitting in Si/Ge superlattices on the order of meV~\cite{chiang1994interference}, which is much smaller than the strain splittings predicted in our study. Therefore, we disregard the effects of intervalley mixing in our analysis. The Si and Ge $\Delta$ valleys correspond to ellipsoidal Fermi surfaces (FS) directed along the [100], [010], and [001] directions, centered at a distance of $\sim0.85(2\pi/a_{\text{Si}})$ and $\sim0.85(2\pi/a_{\text{Ge}})$ from the zone center ($\Gamma-$point) of Si and Ge, respectively. However, we assume that the Si and Ge $\Delta$ valley minima coincide at $(\pm k_{x0},0,0 )$, $(0,\pm k_{y0},0 )$, and $(0,0,\pm k_{z0} )$, with $k_{x0}=k_{y0}=k_{z0}=0.85(2\pi/a_{\text{Si}})$, for the sake of simplicity. Additionally, we consider the mixing of electronic states from pairs of equivalent Si and Ge $\Delta$ valleys that are concentric in the reciprocal space only. It is important to correctly account for the longitudinal ($m_l$) and the transverse ($m_t$) effective masses of the electrons in these $\Delta$ valleys, in order to implement the EMA. The effective masses of the Si and Ge $\Delta$ valleys are very similar to each other with $m_l\sim 0.92m_e$ and $m_t\sim 0.19m_e$~\cite{rieger1993electronic}. We assume that $m_l$ and $m_t$ are constants, independent of strain with $m_l=0.92m_e$ and $m_t=0.19m_e$~\cite{zachai1990photoluminescence}. The dependence of effective masses on strain has been reported to be minimal by a few first principles studies~\cite{yu2008first}. We further ignore the effect of electron-electron interactions in transport. This allows us to assume that the momentum components $(k_x,k_y)$ along in-plane $([100],[010])$ directions of the superlattice is conserved across the interface~\cite{vashaee2004electronic}. However, the cross-plane components of the momentum $k_i$ with $i=$ Si, Ge are not conserved. 

As discussed in the above paragraphs, the superlattice states are formed by the mixing of pairs of equivalent Si and Ge $\Delta$ valleys, concentric in the reciprocal space. These states mix in such a way that the total energy of the electrons $(E)$ remains conserved across the interface, independently, for each pair of the $\Delta_\perp$ and $\Delta_\parallel$ valleys. Splitting $E$ into in-plane ($E_\parallel$) and cross-plane ($E_\perp$) energies, we can write the total energy conservation for electrons from the $\Delta_\perp$ valleys centered at $(0,0,\pm k_{z0} )$ as 
\begin{align}
& E_\parallel + E_\perp \nonumber\\ 
&=\left\{\frac{\hbar^2 (k_x^2+k_y^2)}{2m_t}\right\}+V^\perp_{\text{Si}}+\frac{\hbar^2 (k_\text{Si}\mp k_{z0})^2}{2m_l}\nonumber\\ 
&= \left\{\frac{\hbar^2 (k_x^2+k_y^2)}{2m_t}\right\}+V^\perp_{\text{Ge}}+\frac{\hbar^2 (k_\text{Ge}\mp k_{z0})^2}{2m_l},
\label{eq:crossplane}
\end{align}
and from the in-plane $\Delta_\parallel$ valleys centered at $(\pm k_{x0},0,0 )$ as
\begin{align}
& E_\parallel+E_\perp \nonumber \\
& = \left\{\frac{\hbar^2 (k_x\mp k_{x0})^2}{2m_l}+\frac{\hbar^2 k_y^2}{2m_t}\right\}+V^\parallel_{\text{Si}}+\frac{\hbar^2 k_\text{Si}^2}{2m_t}\nonumber\\ 
&= \left\{\frac{\hbar^2 (k_x\mp k_{x0})^2}{2m_l}+\frac{\hbar^2 k_y^2}{2m_t}\right\}+V^\parallel_{\text{Ge}}+\frac{\hbar^2 k_\text{Ge}^2}{2m_t}.
\label{eq:inplane}
\end{align}
A similar energy balance equation can be written for $\Delta_\parallel$ valleys centered at $(0,\pm k_{y0},0)$ by replacing $k_x$, $k_{x0}, k_y$ in Eq.~\ref{eq:inplane} with $k_y$, $k_{y0}, k_x$, respectively. The terms contributing to $E_\parallel$ are collected in the curly brackets and the rest of the terms contribute to $E_\perp$. It can be clearly seen that the in-plane energy $E_\parallel$ terms within the curly brackets in Eq.~\ref{eq:crossplane} and Eq.~\ref{eq:inplane} are identical in Si and Ge regions. This allows us to solve for the allowed $E_\perp$ in the superlattices corresponding to the six pairs of $\Delta$ valleys, independent of $E_\parallel$. We note that the $\Delta_\perp$ valleys obey $C2$ rotational symmetry about [100] and [010] axes, while the $\Delta_\parallel$ valleys obey $C4$ rotational symmetry about [001] axis. Owing to these symmetry considerations, it suffices to solve for $E_\perp$ corresponding to any one pair of valleys from each type of $\Delta_\parallel$ or $\Delta_\perp$ valleys. Here, we choose to solve for the dispersion relations of the allowed $E_\perp$ states corresponding to the $\Delta_\perp$ valleys centered at $(0,0,k_{z0} )$, and the $\Delta_\parallel$ valleys centered at $(k_{x0},0,0 )$. 

In order to obtain the analytical dispersion relations we analyze the electronic wave functions for the allowed $E_\perp$ states of the superlattice. The in-plane translational symmetry and momentum conservation allow us to separate the wave function envelopes into in-plane $e^{i(k_x x+k_y y)}$ and cross-plane $\psi^j_i(z)$ components, with $i=\text{Si or Ge}$ and $j=$ $\perp$ or $\parallel$, respectively~\cite{trzeciakowski1988effective, vashaee2004electronic}. The EMA or the envelope function approximation further allows us to write $\psi^j_i(z)$ in the Si and Ge regions as a linear combination of their bulk Bloch states. Solving Eq.~\ref{eq:crossplane} for $k_{\text{Si}}$ and  $k_{\text{Ge}}$, we find that the Bloch states at the allowed $E_\perp$ states for the Si and Ge $\Delta_\perp$ valleys correspond to 
\begin{subequations}
\begin{align}
&k_\text{Si}=k_{z0}\pm K \text{ and } k_\text{Ge}=k_{z0}\pm iQ,\text{with} \\
&K=\sqrt{\frac{2 m_l (E_\perp-V^\perp_\text{Si})}{\hbar^2}} \text{ and } Q=\sqrt{\frac{2 m_l (V^\perp_\text{Ge}-E_\perp)}{\hbar^2}},
\end{align}
\label{eq:kBlochperp}%
\end{subequations} 
respectively. Therefore, the cross-plane wave functions from the $\Delta_\perp$ valley states can be written as
\begin{subequations}
\begin{align}
\psi^{\perp}_\text{Si}(z)&=A^{\perp}_\text{Si} e^{i(k_{z0}-K)z}+B^{\perp}_\text{Si} e^{i(k_{z0}+K)z} \\ 
\psi^{\perp}_\text{Ge}(z)&=A^{\perp}_\text{Ge} e^{(i k_{z0}-Q)z}+B^{\perp}_\text{Ge} e^{(i k_{z0}+Q)z}
\end{align}
\label{eq:crossplaneWF}%
\end{subequations}
$\forall$ $E_\perp\geq V^\perp_{Si}$. Similarly, using Eq. (2), the Bloch states at $E_\perp$ for the $\Delta_\parallel$ valleys correspond to
\begin{subequations}
\begin{align}
&k_\text{Si}=\pm K \text{ and } k_\text{Ge}=\pm iQ,\text{with} \\
&K=\sqrt{\frac{2 m_t (E_\perp-V^\parallel_\text{Si})}{\hbar^2}} \text{ and } Q=\sqrt{\frac{2 m_t (V^\parallel_\text{Ge}-E_\perp)}{\hbar^2}},
\end{align}
\label{eq:kBlochpar}%
\end{subequations}
respectively. Therefore, cross-plane wave functions from the $\Delta_\parallel$ valley states can be written as
\begin{subequations}
\begin{align}
\psi^{\parallel}_\text{Si}(z)&=A^{\parallel}_\text{Si} e^{-iKz}+B^{\parallel}_\text{Si} e^{iKz}\\ 
\psi^{\parallel}_\text{Ge}(z)&=A^{\parallel}_\text{Ge} e^{-Qz}+B^{\parallel}_\text{Ge} e^{Qz}
\label{eq:inplaneWF}%
\end{align}
\end{subequations}
$\forall$ $E_\perp\geq V^\parallel_\text{Si}$. The coefficients $A^j_i$ and $B^j_i$ with $i=\text{Si or Ge}$ and $j=$ $\perp$ or $\parallel$, are determined by imposing necessary boundary conditions on $\psi^j_i(z)$. Within the EMA framework, the wave function and its derivative need to obey the Bastard's continuity conditions at the interface\cite{bastard1981superlattice}. Additionally, the wavefunction needs to satisfy the Bloch's condition, yielding $\psi^j_i(z+a)=e^{iqa}\psi^j_i(z)$ and $\psi^j_i(z+a)'=e^{iqa}\psi^j_i(z)'$, for a superlattice with period $a$, and the cross-plane wave vector $q$.  Applying these conditions leads us to the dispersion relations of the $\Delta_\perp$ valley states centered at $(0,0,k_{z0})$ as
\begin{align}
\cos{((q-k_{z0})a)}= \nonumber\\
&\frac{Q^2-K^2}{2K Q}(\sin{(K b)}\sinh{(Q (a-b))})\nonumber\\
&+\cos{(K b)}\cosh{(Q (a-b))},
\label{eq:crossplaneBS}
\end{align}
and of the $\Delta_\parallel$ valley states as 
\begin{align}
\cos{(q a)}= \nonumber\\
&\frac{Q^2-K^2}{2K Q}(\sin{(K b)}\sinh{(Q (a-b))})\nonumber\\&+\cos{(K b)}\cosh{(Q (a-b))}.
\label{eq:inplaneBS}
\end{align}
where $a$ is the superlattice period, $b$ the well width, and $(K,Q)$ determined from Eq.~\ref{eq:kBlochperp} and Eq.~\ref{eq:kBlochpar}, respectively. We solve Eqs.~\ref{eq:crossplaneBS} and~\ref{eq:inplaneBS} numerically to obtain the $E_\perp$ vs $q$ relationship by varying $K$ and $Q$ (as a function of $E_\perp$, Eq.~\ref{eq:kBlochperp}, and Eq.~\ref{eq:kBlochpar}) and solving for $q$. The cross-plane energies $E_\perp$ for $\Delta_\perp$ and $\Delta_\parallel$ valleys thus obtained are superposed with their corresponding $E_\parallel$, shown in Eqs.~\ref{eq:crossplane} and \ref{eq:inplane}, for various $(k_x,k_y)$ to obtain the total dispersion $E$ vs $(q,k_x,k_y)$ across the first Brillouin zone (FBZ).

\subsubsection{Valence Bands}

Thus far we only discussed the dispersion relations of the conduction minibands of Si/Ge superlattices. This is because the electronic transport in $n$-doped Si/Ge superlattices is mainly contributed by the electrons within a narrow region around $E_F$, usually located within the CMB energy window. The superlattices we considered in our study have a high enough band gap such that the valence miniband (VMB) states do not lie this region. As a result, theoretical predictions of electronic properties in $n$-doped Si/Ge superlattices included only the CMB and ignored the contribution from the VMB~\cite{koga1999carrier}. Nevertheless, for the sake of presenting a complete analysis of the electronic bands of Si/Ge superlattices using EMA, we briefly discuss the derivation of VMB dispersion relations considering a simplified model for the anisotropic heavy holes (HH) and light holes (LH) of Si and Ge. Similar to the $\Delta$ valley case, it is important to correctly account for the potential profiles and the effective masses of the HH and LH in the Si and Ge regions of the superlattice. The valence band maxima (VBM) of LH and HH coincide at the $\Gamma$ point for unstrained Si and Ge and undergo degeneracy lifting under strain. We compute the strain controlled LH and HH potential profiles of Si and Ge in a [001] Si/Ge superlattice using the deformation potential theory~\cite{van1986theoretical}. We denote the strain-split HH and LH VBM with $V^{j}_{i}$ with $i=$ Si or Ge and $j=$ HH or LH, respectively. The FS of HH and LH are ellipsoids centered at the $\Gamma$ point with longitudinal and transverse effective masses that have nonlinear strain dependence\cite{chun1992effective,yu2008first,sun2007strain}. However, for the sake of simplicity, we assume that the HH effective masses are strain independent with longitudinal effective masses $m^{HH}_{l,\text{Si}}$=0.28$m_0$ and $m^{HH}_{l,\text{Ge}}$=0.22$m_0$ and transverse effective masses $m^{HH}_{t,\text{Si}}$=0.22$m_0$ and $m^{HH}_{t,\text{Ge}}$=0.06$m_0$. On the other hand, we assume the LH longitudinal effective masses linearly decrease from 0.20$m_0$ to 0.18$m_0$ for Si ($m^{LH}_{l,\text{Si}}$) and increase from 0.05$m_0$ to 0.14$m_0$ for Ge ($m^{LH}_{l,\text{Ge}}$)~\cite{zachai1990photoluminescence}, due to the lowest to highest strain experienced by Si (tension) and Ge (compression), respectively. We assume the LH transverse effective masses are strain independent at $m^{LH}_{t,\text{Si}}$=0.25$m_0$ and $m^{LH}_{t,\text{Ge}}$=0.07$m_0$. 

With the effective masses characterized, we can write the energy conservation of the HH states across the interface as
\begin{align}
E & = V^{HH}_\text{Si}+\frac{\hbar^2 k_\text{Si}^2}{2m^{HH}_{l,\text{Si}}}+\left\{\frac{\hbar^2 (k_x^2+k_y^2)}{2m^{HH}_{t,\text{Si}}}\right\}\nonumber\\ 
&= V^{HH}_\text{Ge}+\frac{\hbar^2 k_\text{Ge}^2}{2m^{HH}_{l,\text{Ge}}}+\left\{\frac{\hbar^2 (k_x^2+k_y^2)}{2m^{HH}_{t,\text{Ge}}}\right\}.
\label{eq:crossplaneV}
\end{align}
A similar relation can be written for the LH state replacing the HH effective masses with that of the LH in Eq.~\ref{eq:crossplaneV}. We note that the in-plane energy components in Si and Ge regions shown in the curly brackets in Eq.~\ref{eq:crossplaneV} are not similar to each other unlike the $\Delta$ valley case, due to the difference in the transverse effective masses. Therefore, the total energy $E$ cannot be split into $E_\perp$ and $E_\parallel$ components. We solve for the allowed $E$ vs $(q,k_x,k_y)$ employing EMA for various total in-plane momenta $k_{||}=\sqrt{k_x^2+k_y^2}$~\cite{bastard1981superlattice,mukherji1975band}. Here, the $E$ at a given $q$ and $k_\parallel$ corresponds to all possible ($k_x$,$k_y$) pairs that satisfy the chosen $k_\parallel$. By choosing various $k_\parallel$, we compute the VMB across the FBZ in a manner similar to the CMB. For a rigorous calculation of the VMB including a full consideration of the anisotropy and the strain-dependence of the hole effective masses, the reader is advised to consult other references~\cite{chun1992effective,yu2008first,sun2007strain}. These studies provide particularly useful insight to analyze electronic transport in $p$-doped Si/Ge superlattices employing the EMA. 

\subsection{First-Principles Energy Dispersion Relations of Si/Ge Superlattice Bands} 
In order to establish the reliability of transport property predictions using EMA, it is important to understand how the EMA bands compare to those obtained with a higher-accuracy numerical method. We compare the EMA bands of strained Si$_4$Ge$_4$ superlattices with those obtained with DFT (Fig.~\ref{fig:EMADFT}). The electronic structure properties are obtained using the plane-waves code Quantum Espresso (QE)~\cite{giannozzi2009quantum}. Our Si$_4$Ge$_4$ model superlattice supercell consists of 4 and 8 monolayers (MLs) in the in-plane and cross-plane directions, respectively, and corresponds to a tetragonal Brillouin zone. To simulate the effect of substrate strain, we fix the in-plane lattice constant $a_\parallel$ corresponding to the substrate lattice constant, inducing the in-plane strain  $\epsilon_{\text{Si},\parallel}$. We then relax the  superlattice in the [001] cross-plane direction by performing a self-consistent calculation (SCF) using the Broyden-Fletcher-Goldfarb-Shanno Quasi-Newton algorithm. The SCF calculations are performed on a $4\times4\times2$ $k$-mesh using the generalized gradient approximation (GGA) of the Pedrew-Burke-Ernzerhof (PBE) exchange-correlation functional by employing scalar relativistic normconserving pseudopotentials for
Si and Ge atoms~\cite{perdew1996generalized}. We generate the $k$-mesh using the
Monkhorst-Pack scheme which significantly reduces the computational time owing to the superlattice symmetry~\cite{monkhorst1976special}. An energy cutoff of 30 Ry was used to expand the Kohn-Sham orbitals in terms of a plane wave basis set~\cite{van1986theoretical,satpathy1988electronic}. A convergence threshold of $10^{-9}$ Ry was used for self-consistency. Following the SCF calculations, we perform the electronic structure calculations for the Si$_4$Ge$_4$ superlattice using non self-consistent field (NSCF) calculations. We use a dense $k$-mesh of $40\times40\times20$ for all our NSCF calculations. We ignore the spin-orbit (SO) coupling effects in our analysis since it was shown that the strain splittings of the bands is considerably large compared to the SO splittings~\cite{hybertsen1987theory}. 
\subsection{Electronic Transport Coefficients Using Boltzmann Transport Equation}

We use BTE with the constant relaxation time approximation to compute the electronic transport properties of $n$-doped Si/Ge superlattices. Specifically, the cross-plane Seebeck coefficient $S$, and the electrical conductivity $\sigma$ are obtained using the following expression~\cite{mahan1996best,ashcroftmermin}:
\begin{align}
\mathcal{L}^{(a)} &= \tau \int dE \left[v_{g}^2 \rho_{DOS}(E)(E-E_F)^a \left(-\frac{\partial f_0(E)}{\partial E}\right) \right] \label{eq:L_integral}\\
\sigma &=\mathcal{L}^{(0)}
\label{eq:conductivity}\\
S &= \frac{1}{eT} \frac{\mathcal{L}^{(1)}}{\mathcal{L}^{(0)}} \label{eq:seebeck}\
\end{align}
where $e$ is electron charge, $T$ is temperature, $\tau$ is the electron relaxation time, $E$ is energy, $v_{g}(E)$ is the average cross-plane group velocity of an electron with energy $E$, $\rho_{DOS}(E)$ is superlattice density of states (DOS), $E_F$ is the Fermi energy, and $f_{0}(E)$ is the Fermi-Dirac distribution function.  

\section{Results and Discussion}
\begin{figure}[tp]
\centering
\includegraphics[width=1.0\linewidth]{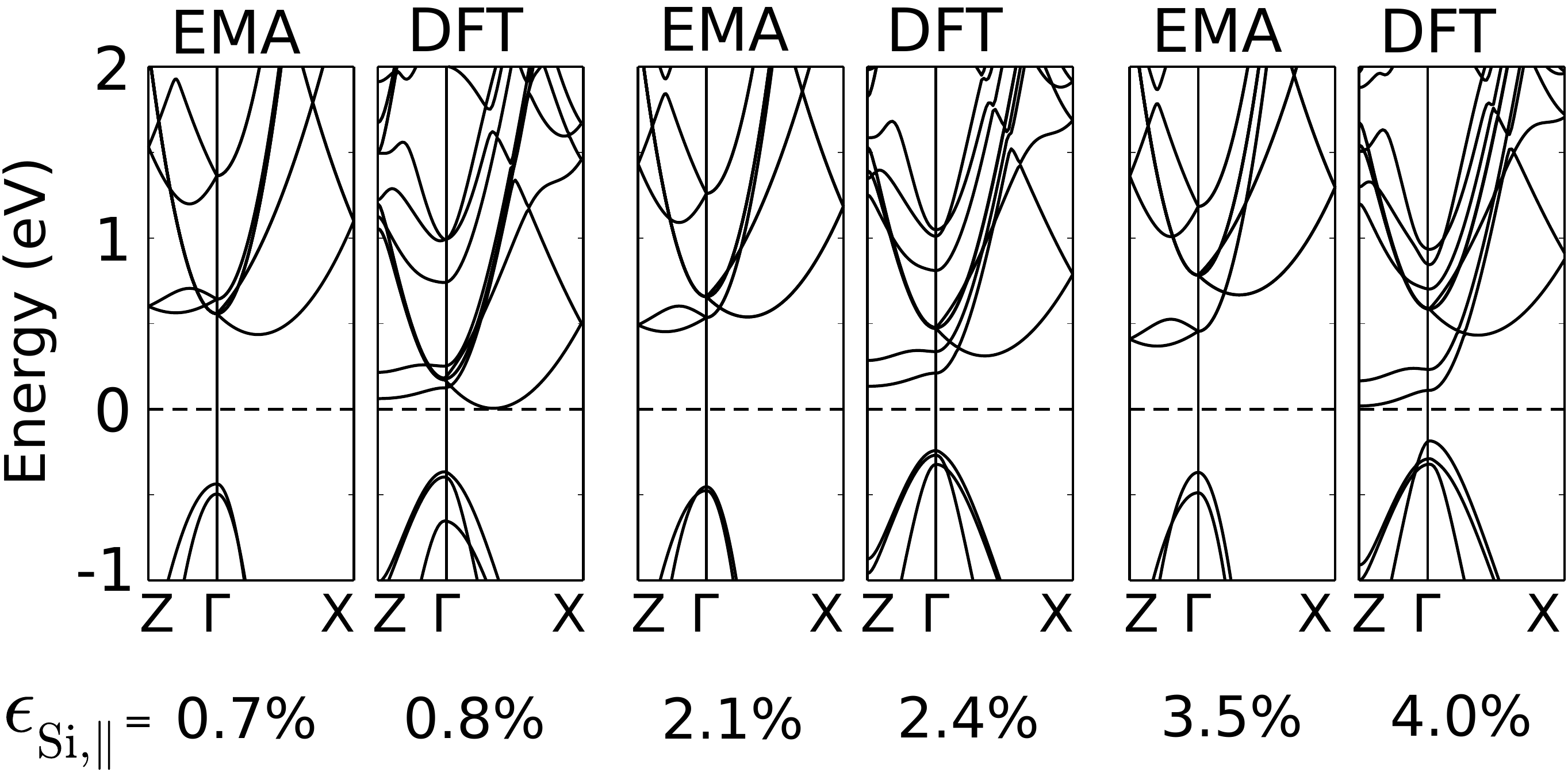}
\caption{Comparison of electronic bands of substrate strained Si$_4$Ge$_4$ superlattices obtained using DFT and EMA. We consider the superlattices with the substrate induced strain values shown at the bottom of the figure.}
\label{fig:EMADFT}
\end{figure}

We now discuss our main results demonstrating various band engineering approaches to modulate the thermopower and other electronic transport coefficients of $n$-type Si/Ge superlattices, using EMA. 

\subsection{Comparison Between Energy Bands Obtained with Different Methods}
We show the comparison between the energy bands of substrate strained Si$_4$Ge$_4$ superlattices computed with EMA and DFT in Fig.~\ref{fig:EMADFT}, to establish the predictive power of the EMA. The respective substrate induced in-plane strain values are shown at the bottom of the figure. The EMA models are chosen to represent superlattices with $\epsilon_{\text{Si},\parallel}=\{0.7\%, 2.1\%, 3.5\%\}$. We compare the band structures of these EMA models with the DFT models representing $\epsilon_{\text{Si},\parallel}=\{0.8\%, 2.4\%, 4.0\%\}$. As can be noted, the strain values as well as the lattice parameters do not have a one-to-one comparison between the two methods. The unstrained Si and Ge lattice constants predicted by DFT and EMA are $a^{\text{DFT}}_{\text{Si}}=5.475\;\text{\AA}$, $a^{\text{DFT}}_{\text{Ge}}=5.740\;\text{\AA}$ and $a^{\text{EMA}}_{\text{Si}}=5.431\;\text{\AA}$, $a^{\text{EMA}}_{\text{Ge}}=5.658\;\text{\AA}$, respectively. The in-plane strain in confined Si, $\epsilon_{\text{Si},\parallel}$, of the superlattice DFT models grown on a Si or a Ge substrate, ranges from 0\%-4.8\%, respectively. This is in contrast with the respective 0\%-4.2\% range present in the EMA models. To establish a common reference for comparison, we scale the EMA predicted $\epsilon_{\text{Si},\parallel}$ in the 0\%-4.2\% range to fit within the DFT predicted 0\%-4.8\% range, and estimate the strain in an equivalent DFT model. We further align the Fermi levels ($E_F$) of the EMA and DFT bands and set them to 0 eV, to facilitate the band structure comparison. We note that the splitting of $\Delta$ valleys and their strain induced relative movements are very well predicted by both the DFT and the EMA approaches. However, EMA models consider a larger band gap compared to DFT which leads to a misalignment of the minibands. This happens because of the systematic underestimation of the band gaps by the Kohn-Sham states in the DFT~\cite{perdew1986density}. The PBE functional, in particular, predicts an incorrect 0 eV band gap for Ge~\cite{heyd2005energy}. While in the EMA, we correctly account for the band gaps of unstrained Si and Ge by matching them with their experimental values of $\sim 1.17$ eV and $\sim 0.96$ eV, respectively~\cite{van1986theoretical, lang1985measurement}. Therefore, one needs to adjust the DFT bands shown in Fig.~\ref{fig:EMADFT} with the correct band gaps to show a better overall match between the EMA and DFT band structures. Nevertheless, the EMA and the DFT CMB match well, except that EMA does not predict sub-band splittings due to excluding inter-valley mixing effects. With the reliability of the EMA bands established by comparing with DFT, we proceed to analyze the electronic transport properties of the superlattices using these bands, which we perform at $T=300$ K unless mentioned otherwise.
\begin{figure*}[t]
\centering
\includegraphics[width=1.0\linewidth]{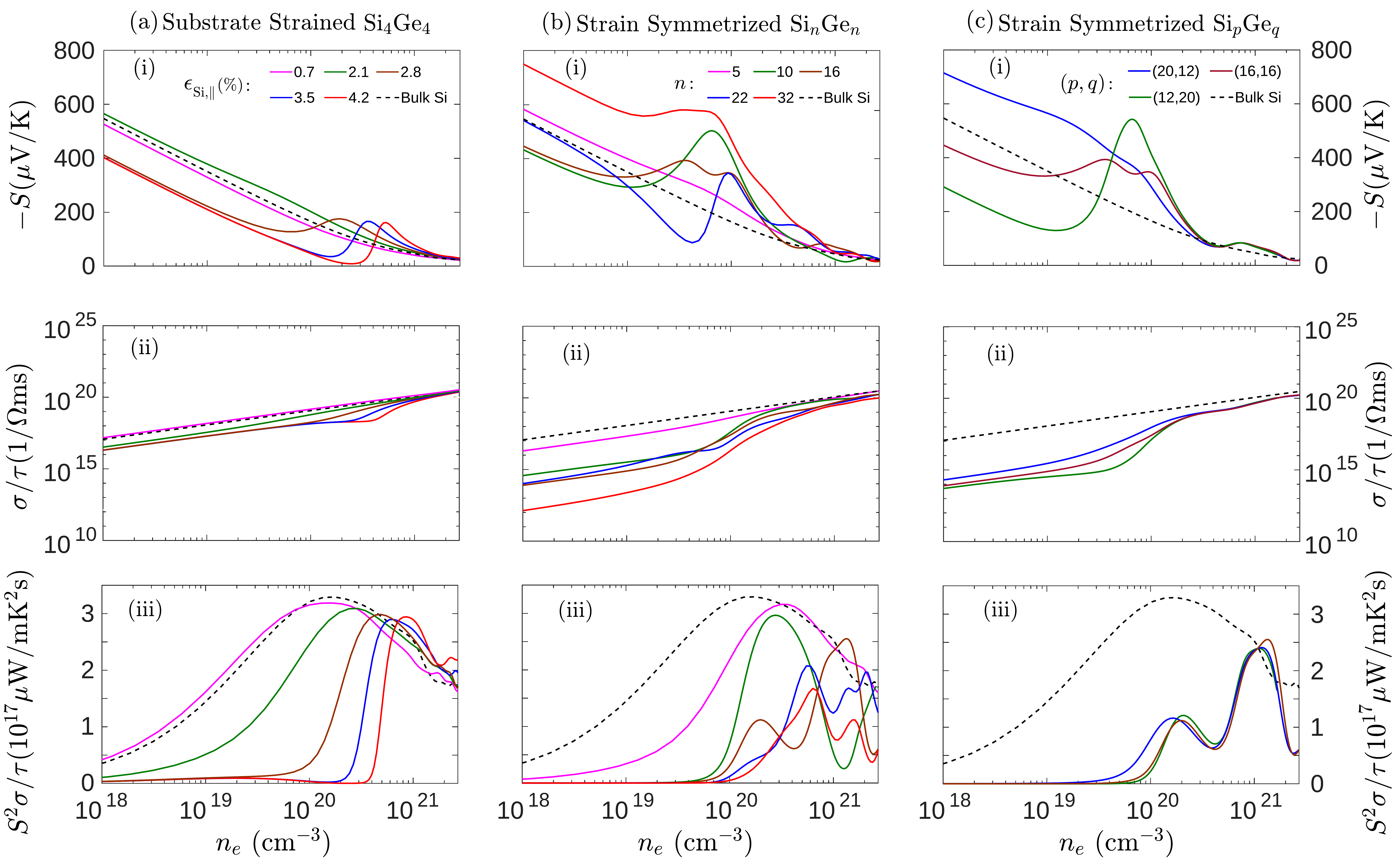}
\caption{Electronic transport properties of Si/Ge superlattice obtained with EMA: (a) substrate strained Si$_4$Ge$_4$ superlattices, (b) Si$_n$Ge$_n$ superlattices with varying periods, (c) Si$_p$Ge$_q$ superlattices with a fixed period of $p+q=32$ and varied well (Si) and barrier (Ge) widths. The top panels display the thermopower or Seebeck coefficients ($S$), the middle panels display electronic conductivity ($\sigma$) and the bottom panels display the electronic power factors ($S^2\sigma$), in units of $\tau$, respectively. All calculations are performed at 300 K.}
\label{fig:Strain}
\end{figure*}

\subsection{Electronic Transport Coefficients of Si/Ge Superlattices}

\subsubsection{Substrate Strained Superlattices}

We demonstrated that strain induced CMB modulations enable tunable enhancements of the electronic thermopower of the Si/Ge superlattices, especially in the high-doping regime, in our recent work using DFT~\cite{proshchenko2019optimization,proshchenko2019modulation}. Here, we aim to illustrate that such strain-controlled modulations of $S$ are also captured within a simpler EMA approach, as this further substantiates our EMA methodology and helps us to form physical intuition about the system. To this end, we compute the electronic transport coefficients of the strained Si$_4$Ge$_4$ superlattices with $\epsilon_{\text{Si},\parallel}$=\{0.7\%, 2.1\%, 2.8\%, 3.5\%, 4.2\%\} employing the EMA. We present the EMA-BTE predicted $S$ of the strained superlattices and that of bulk Si in Fig.~\ref{fig:Strain}(a)(i). The $S$ of bulk Si monotonically decreases with increasing carrier concentration ($n_e$), following a Pisarenko-like relation (PR)~\cite{hinsche2012thermoelectric}. The $|S|$ of strained Si$_4$Ge$_4$ superlattices shows a bulk-like monotonic behavior in the low-strain regime ($\epsilon_{\text{Si},\parallel}=0.7$\%), and is reduced compared to bulk Si. As we increase $\epsilon_{\text{Si},\parallel}$, we note the emergence of an oscillatory behavior as a function of the carrier concentration $n_e$, breaking the PR. A general trend can be noted that the peaks move towards higher $n_e$ with increasing strain. A similar trend was observed in the first principles DFT study as well~\cite{proshchenko2019modulation}. Our EMA-BTE approach predicts a $\sim2.4$-fold enhancement of $S$ at $n_e=5.2\times10^{20}$ cm$^{-3}$ for a substrate strained Si$_4$Ge$_4$ superlattice with $\epsilon_{\text{Si},\parallel}$=4.2\%. While enhancement of $S$ has its own advantages, it is important to characterize how the strain induced CMB modulations affect other electronic transport coefficients, to use the strain engineering approach for a broad range of technological applications including thermoelectric applications. 

We present the variation of $\sigma$ and PF (in the units of $\tau$) of strained Si$_{4}$Ge$_{4}$ superlattices with $n_e$ in Figs.~\ref{fig:Strain} (a)(ii) and (iii), respectively. The $\sigma/\tau$ of bulk Si monotonically increases with increasing $n_e$ compensating the monotonic decrease of the $S$. This $S-\sigma$ trade-off results in the PF peak at $n_e\sim3.2\times10^{20}$ cm$^{-3}$, in a similar manner discussed in previous studies~\cite{hinsche2012thermoelectric}. Interestingly, $\sigma$ of Si$_4$Ge$_4$ superlattices are slightly increased in the low strain (0.7\%) regime, compensating for the decrease in $S$. As a result, the PF peak shifts towards lower $n_e$ resulting in a $\sim 10\%$ increase of PF for $n_e<5\times10^{19}$ cm$^{-3}$. However, $\sigma$ is reduced in the superlattices with moderate (2.1\%) to high (4.2\%) substrate strain. This results in a reduced PF for $n_e<10^{20}$ cm$^{-3}$ for moderate to high strain cases. A trend can be noted that the PF peak becomes sharper and shifts towards higher $n_e$ with increasing strain. PF is $\sim20\%$ improved at $n_e\sim11\times10^{20}$ cm$^{-3}$ for the highest strain case (4.2\%). This re-establishes the idea that the miniband modifications induced by the substrate strain can help modulate electronic transport in superlattices. 

\subsubsection{Strain-Symmetrized Superlattices with Varied Period}

However, it was found that it is energetically unfavourable to grow strained Si/Ge superlattices~\cite{kasper1988symmetrically,kasper1989strain,kasper1986layered,pearsall1987structurally}. Instead, SS superlattices are more stable and can be grown easily~\cite{kasper1986layered}. In these superlattices, strain originates due to the lattice mismatch between the superlattice components. Hence, it is of practical interest to explore the strain induced electronic properties of SS superlattices. Previous studies reported that the electronic transport in the SS superlattices can be modulated by varying the superlattice periods and the layer thicknesses~\cite{koga2000carrier,bahk2012seebeck, hinsche2012thermoelectric,bottner2006aspects,kasper1988symmetrically}. In Fig.~\ref{fig:Strain} (b), we present the electronic properties of SS Si$_n$Ge$_n$ superlattices with varied periods, $L=2n$, where $n=\{5,10,16,22,32\}$ MLs. Interestingly, the in-plane strain in the confined silicon components of the SS superlattices is a constant, $\epsilon_{\text{Si},\parallel}$=1.9\%, independent of $n$~\cite{van1986theoretical}. Since the $\epsilon_{\text{Si},\parallel}$ is the same for all the chosen superlattices, the potential profiles in Si and Ge regions are also the same. The only variables are the widths of the Si (well) and Ge (barrier) regions, which vary by the same number of MLs, with the variation of the period of the superlattice. We predict that these superlattices display an oscillatory $S$ with respect to $n_e$ as shown in Fig. \ref{fig:Strain}(b)(i), breaking the PR. Particularly, the $n=10$, 16, and 22 superlattices show a strong oscillatory behavior with regions of increase and decrease of $S$. While the $n=5$ and 32 superlattices show an increased $S$ for all $n_e$ considered, with the $n=32$ superlattice showing a remarkable increase of overall $S$. The thermopower of the $n=32$ superlattice shows a maximum $\sim 3.2$-fold enhancement at $n_e=7\times10^{19} $cm$^{-3}$. However, the increase of $S$ is compensated by decrease of $\sigma$ as shown in Fig.~\ref{fig:Strain}(b) (ii), in a similar manner to that of the substrate strained superlattices discussed before. This $S-\sigma$ trade-off results in the drastically diminishing PF for $n_e<10^{20} $cm$^{-3}$ as shown in Fig. \ref{fig:Strain}(b) (iii). On the other hand, in the high-doping regime $n_e>5\times10^{20} $cm$^{-3}$, we observe a 10-20$\%$ enhancement of PF for $n=5$, 16, and 22 superlattices. Through this analysis, we establish that modulations in the electronic transport properties can be achieved by varying period of the SS Si/Ge superlattices.

\subsubsection{Strain-Symmetrized Superlattices with Fixed Period and Varied Compositions}

The strain-symmetrized superlattices we have discussed thus far have a constant potential profile owing to the equal number Si and Ge MLs. However, the strain in the SS superlattices vary when the number of Si ($p$) and Ge MLs ($q$) are varied independently~\cite{van1986theoretical}. The non-uniform strain leads to variable potential profiles in the Si and Ge regions. Additionally, the ratio of well to barrier width is not constant in these cases. Here we investigate Si$_p$Ge$_q$ superlattices with $p+q=32$ MLs. In Fig.~\ref{fig:Strain} (c), we present the electronic transport properties of Si$_p$Ge$_q$ superlattices with $(p,q)$=\{(20,12), (16,16), (12,20)\} yielding symmetrized strains, $\epsilon_{\text{Si},\parallel}$=\{1.4\%, 1.9\%, 2.4\%\}. In Fig.~\ref{fig:Strain} (c)(i), we find that for $p>q$, $S$ in the low-doping regime is considerably enhanced while maintaining an overall improvement in the mid to high doping regimes. Increasing the Ge MLs tends to push the high-$S$ regions to higher $n_e$ preceded by low regions in the form of oscillatory peaks. The Si$_{12}$Ge$_{20}$ superlattice shows a significant $\sim3$-fold $S$ enhancement compared to the bulk Si at $n_e=7\times10^{19} $cm$^{-3}$. Interestingly, among the three cases studied, the $S-\sigma$ trade-off plays in such a way that the $p=q$ case gives the maximum enhancement in the PF in the high-doping regime.

\subsubsection{Qualitative Explanation of the Modulation of Electronic Transport in Superlattices}

We demonstrated that the Pisarenko-like $S$ vs $n_e$ inverse relationship can be broken in substrate strained and strain-symmetrized $n$-doped Si/Ge superlattices. Only a few studies reported such a behavior in Si/Ge superlattices~\cite{hinsche2012thermoelectric,proshchenko2019optimization,proshchenko2019modulation}. It is therefore important to understand the mechanism that governs this behavior to aid future research. At a given temperature, $S \propto \mathcal{L}^{(1)}/\sigma$ in the BTE framework (Eq.~\ref{eq:seebeck}). Therefore, to understand the behavior of $S$ with $n_e$, it is imperative to understand the functional relationship of $\mathcal{L}^{(1)}$ and $\sigma$ with $n_e$. We note that {\em both} $\mathcal{L}^{(1)}$ and $\sigma$ are determined from the integrals containing the product $v_{g}^2\rho_{DOS}\left(-\frac{\partial f_0}{\partial E}\right)$. The term $v_{g}^2\rho_{DOS}$ can be thought of as the electronic contribution to transport at energy $E$. While the term $\left(-\frac{\partial f_0}{\partial E}\right)$, referred to as the Fermi window (FW), determines the energy window in which the dominant electronic contribution to transport occurs. The FW is centered at and symmetric with respect to $E_F$ and  full-width at half-maximum $\sim3.5k_BT$, where $k_B$ is the Boltzmann constant. As the $n_e$ is increased at a fixed $T$, the FW shifts in energy as $E_F$ raises in energy. The $\mathcal{L}^{(1)}$ integral {\em additionally} includes the product of $(E-E_F)$ and the FW, resulting in an anti-symmetric window (ASW) function in the numerator. Due to the presence of the ASW in $\mathcal{L}^{(1)}$, the electrons with energy above $E_F$ contribute \emph{positively} towards the net $S$ and are referred to as hot electrons. While, the electrons with energy below $E_F$ contribute \emph{negatively} towards $S$ and are referred to as cold electrons. This distinction between the integrands of $\mathcal{L}^{(1)}$ and $\sigma$ can be used to explain the intriguing behavior of $S$ and the $S-\sigma$ trade-off mentioned before. We present a brief discussion of this aspect in the following paragraphs to explain our observations. The interested reader is encouraged to consult previous studies to acquire more understanding of the physical phenomena~\cite{vashaee2004electronic,bahk2012seebeck,vashaee2006cross, vashaee2007thermionic}. 

$\rho_{DOS}$ and $v_{g}$ monotonically increase with $E$ in bulk Si owing to the parabolic approximation of $\Delta$ valleys made in the EMA. Therefore as $n_e$ is increased, more electronic states fall within FW. With this understanding, it naturally follows from Eq.~\ref{eq:conductivity} that $\sigma$ of bulk Si monotonically increases with $n_e$. On the other hand, due to the presence of ASW in $\mathcal{L}^{(1)}$, we find that $\sigma/\mathcal{L}^{(1)}$ also monotonically increases with $n_e$. This happens because $\mathcal{L}^{(1)}$ increases at a lower rate with $n_e$ compared to $\sigma$. This explains the Pisarenko-like $S$ vs $n_e$ inverse relationship observed for bulk Si. In contrast, the formation of minibands and the degeneracy splitting of $\Delta$ valleys in [001] Si/Ge superlattices lead to non-monotonic step-like $\rho_{DOS}$ and $v_{g}$~\cite{prairie1990general,koga1999carrier}. This results in the oscillatory $S$ shown in Fig. \ref{fig:Strain}. $S$ is increased when the rate of increase of $\mathcal{L}^{(1)}$ with $n_e$ is greater than that of $\sigma$ with $n_e$. This happens at an $n_e$ when the corresponding $E_F$ approaches a miniband edge resulting in a sharp increase of the $v_{g}^2\rho_{DOS}$ above the $E_F$ leading to a transport that is strongly in favour of hot electrons over the cold electrons. We elucidate this direct relationship between $S$ and $v_{g}^2\rho_{DOS}$ below to explain the electronic transport properties of the superlattices shown in Fig.~\ref{fig:Strain} in columns (a), (b), and (c). 

(a) In strained Si$_4$Ge$_4$ superlattices, the minibands from the $\Delta_\parallel$ valleys move upwards in energy with respect to those from the $\Delta_\perp$ valleys, with increasing strain. This energy shift creates oscillatory peak in the $S$ vs $n_e$ curve that moves towards higher $n_e$ as $\epsilon_{\text{Si},\parallel}$ increases from 2.1\% to 4.2\% as shown in Fig.~\ref{fig:Strain} (a). The high $n_e$ peak also narrows with increasing strain since the first $\Delta_\parallel$ miniband states move closer to the second $\Delta_\perp$ miniband states as shown in Fig.~\ref{fig:EMADFT}. 

(b) In strain-symmetrized Si$_n$Ge$_n$ superlattices with varied periods, we increase the well and barrier widths simultaneously. As explained before, this implies the potential profile of the $\Delta_\perp$ and $\Delta_\parallel$ valleys within the well and barrier regions is unaffected. Therefore, the modulation of the minibands in these superlattices is primarily due to the reduction of the BZ in the cross-plane direction. In a reduced BZ, the number of minibands is increased owing to the band folding effects. Therefore, the FW and ASW \emph{include} more oscillations of the $v_{g}^2\rho_{DOS}$ which manifests as a strong oscillatory $S$, especially for $n=10$, 16, and 22 superlattices. Interestingly, the oscillatory nature is less prominent in the $n=32$ superlattice. This is because although there are more minibands in the $n=32$ superlattice, some minibands are sufficiently close to each other that the FW and ASW cannot \emph{distinguish} them as distinct bands.

(c) In strain-symmetrized Si$_p$Ge$_q$ superlattices with fixed periods, we vary the composition of the well and the barrier components keeping the total number of the MLs fixed. In all the superlattices, the period and hence the superlattice BZ is approximately invariant. Therefore, the major factors that influence the miniband dispersion are the variable symmetrized strains, and the well and the barrier widths. A variable symmetrized strain, as explained previously, implies a varying potential profile in the well and the barrier regions. When $p>q$, the $a_\parallel$ of the superlattice tends to relax close to $a_\text{Si}$ and the potential profiles tend towards that of the low substrate strain case. This effect, in conjunction with the increased well width, results in a sharper increase of the $v_{g}^2\rho_{DOS}$ in the low-doping regime compared to bulk Si. This in turn leads to a considerable improvement of $S$ in the low-doping regime. On the other hand, in the case of $p<q$, the $a_\parallel$ of the superlattice tends to relax close to $a_\text{Ge}$ and the potential profiles tend towards that of the high substrate strain case. Moreover, the barrier width is considerably increased in this case which further narrows the minibands. These effects together lead to a sharp increase of the $v_{g}^2\rho_{DOS}$ leading to a local $S$ peak in the high-doping regime. 

\begin{figure}
\centering
\includegraphics[width=1.0\linewidth]{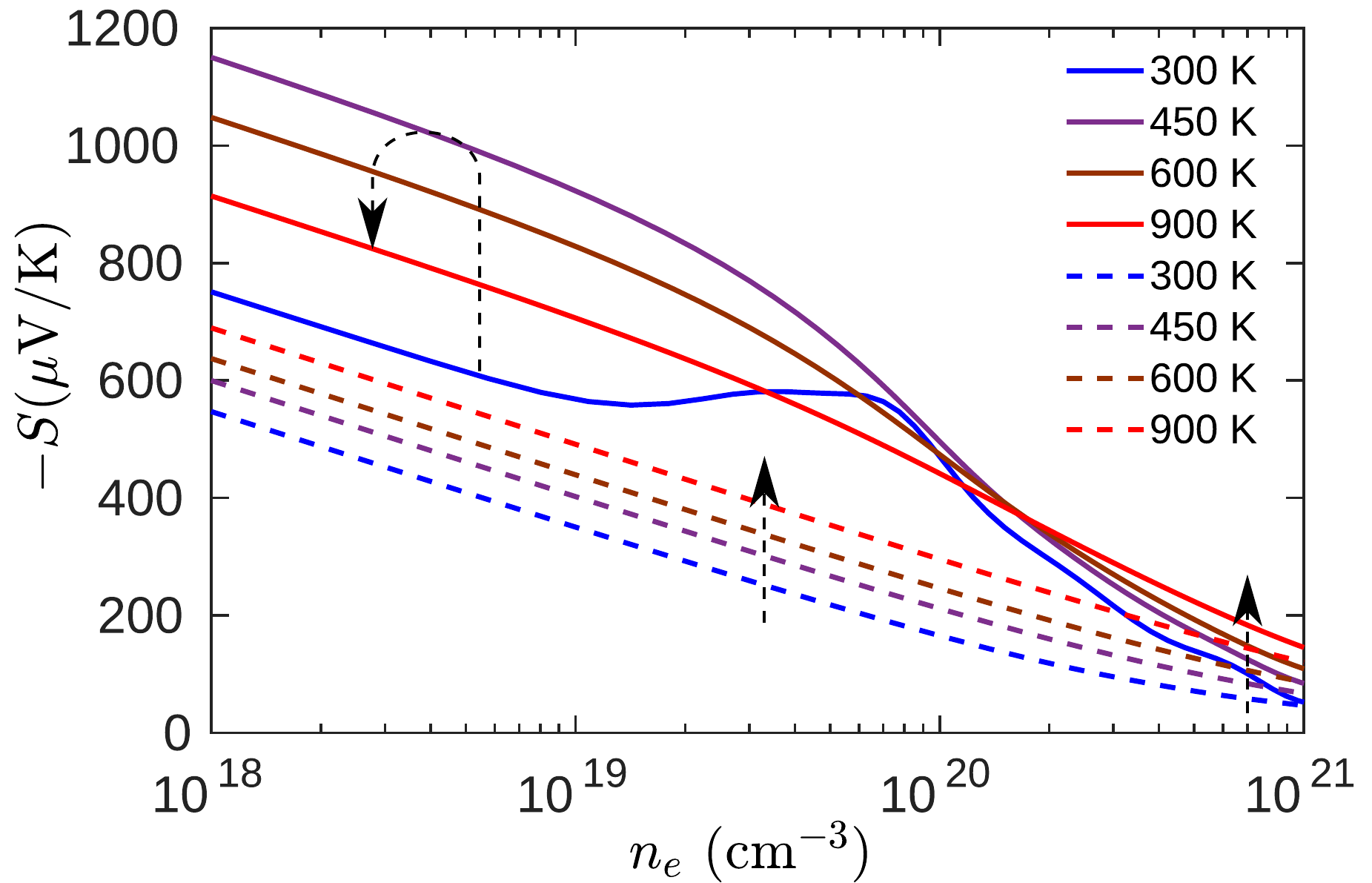}
\caption{Temperature dependent $S$ of Si$_{32}$Ge$_{32}$ superlattice (solid lines) and bulk Si (dashed lines) as a function of carrier concentrations.}
\label{fig:SvsT}
\end{figure}

\subsubsection{Effect of Temperature on Thermopower of Superlattices}

We have demonstrated the various ways to modulate $S$ as a function of $n_e$ at a fixed $T=300$ K, by modifying the superlattice band structure with varied strain, period, and composition. However, the FW width varies with $T$ resulting in a temperature dependence in $S$. In degenerate semiconductors, there exists a direct relationship between $S$ and $T$ at a fixed doping level~\cite{snyder2011complex}. In Fig.~\ref{fig:SvsT}, we show a monotonically increasing $S-T$ relationship for bulk Si (dashed lines) $\forall$ $n_e$ considered. However, we find that this monotonic relationship is broken in Si/Ge superlattices, especially at the technologically relevant doping regime. $S$ of Si$_{32}$Ge$_{32}$ superlattices increases with $T$ up to 450 K and drops as we go higher in $T$, for $n_e<10^{19} $cm$^{-3}$, as shown in Fig.~\ref{fig:SvsT} (solid lines). This can be qualitatively explained from the understanding that the FW and hence the ASW broaden with increasing $T$ leading to a variation in the contribution from the hot and the cold electrons to $S$ with $T$ at a given $n_e$. In the case of bulk Si, the ASW broadening leads to an increased contribution from the hot electrons due to a monotonically increasing $v_{g}^2\rho_{DOS}$. While the non-monotonic nature of the $v_{g}^2\rho_{DOS}$ results in the observed behavior in the Si$_{32}$Ge$_{32}$ superlattice. As $T$ is increased to 450K, the superlattice miniband states that resulted in a bump in $S$ at $T=300$ K around $n_e\sim7\times10^{19} $cm$^{-3}$ (Fig.~\ref{fig:Strain}), contribute to $S$ at a lower $n_e$ due to the ASW broadening. Therefore, we see up to $\sim2.4$-fold $S$ enhancement at $T=450$ K in the low-doping regime. However, further increase of $T$ extends the ASW to include the shallow $v_{g}^2\rho_{DOS}$ region, corresponding to the cross-plane miniband gap that follows the miniband states that led to the bump at 300 K, resulting in a decreasing $S$. In the high-doping regime, $n_e>3\times10^{20} $cm$^{-3}$, $S$ monotonically increases with $T$ approaching a Pisarenko-like behavior as observed in bulk Si. The $E_F$ is high enough that the miniband like nature is less apparent in the high-doping regime.

\section{Summary and Outlook}
We analyzed the cross-plane miniband transport in $n$-doped [001] Si/Ge superlattices with the effective mass approximation, and explored ways to enhance the electronic thermopower and power factor. To the best of our knowledge, only direct-gap based EMA has been employed so far to investigate the electronic transport properties of the Si/Ge superlattices. Here we established a new indirect-gap based EMA approach to correctly account for the indirect nature of the Si and Ge band gaps in the analysis. We compared the energy bands obtained with EMA with those computed with DFT to discuss the reliability of the approach. Using BTE framework in combination with the EMA band analysis, we uncovered that $S$ of $n$-doped Si/Ge superlattices can be enhanced up to $\sim 3.2$-fold in high doping regimes, breaking the Pisarenko relation. We demonstrated that this tunability can be achieved by growing the superlattices on various substrates, and varying superlattice period, and the composition. The increase of $S$ is largely compensated by the decrease in $\sigma$ leading to a reduced PF in most of these cases. However, we observed  modest improvement of the PF of superlattices under a low (high) substrate strain, in low (high) doping regimes. We note improvements of the PF of symmetrically strained superlattices in the high-doping regimes as well with varying period and composition. We show that in addition to varying non-monotonically with increasing $n_e$ due to lattice strain, $S$ shows a non-monotonic increase with increasing $T$ as well, especially in the low-doping regime. Therefore, further analysis is required to estimate the electronic transport properties of [001] Si/Ge superlattices at varied strain environments or desired temperatures. Further improvements could be made in the current work by including the inter-valley mixing effects, non-conservation of transverse momentum, and a non-constant electronic relaxation time. In addition, perturbation theory may be employed in conjunction to correctly account for the band splittings at the Brillouin zone boundaries~\cite{misra2011physics}. We hope that our analyses act as preliminary predictions encouraging further theoretical and experimental research to validate our findings. Our approach can be extended to other superlattice systems as well. For instance, one can use the methods presented in this work to study dimensional effects on electronic transport in two-dimensional nanowire superlattices and three-dimensional nanodot superlattices. Numerical matrix method based approaches may be employed to carry on the analysis~\cite{le2016numerical,pavelich2015kronig,pavelich2016calculation}. We anticipate that the ideas presented here will have a strong impact in controlling electronic transport in various thermoelectric, opto-electronic, and quantum-enhanced materials applications. 

\section{Acknowledgements}
The work is funded by the Defense Advanced Research Projects Agency (Defense Sciences Office) [Agreement No.: HR0011-16-2-0043]. All computations were performed using the Extreme Science and Engineering Discovery Environment (XSEDE), which is supported by National Science Foundation grant number ACI-1548562.

\bibliography{references}

\end{document}